\documentclass{article}
\usepackage[T1]{fontenc}
\usepackage{iclr2027_conference,times}
\usepackage{amsmath,amssymb,amsfonts}
\usepackage{booktabs,array,graphicx}
\usepackage{xcolor,tikz}
\usetikzlibrary{arrows.meta,positioning,fit,calc,backgrounds}
\usepackage{xurl}
\usepackage[hidelinks,breaklinks=true]{hyperref}
\newcommand{\E}{\mathbb{E}}
\newcommand{\R}{\mathbb{R}}
\newcommand{\omctablefont}{\footnotesize}
\newcommand{\authorblock}{%
\small
\begin{tabular}{@{}l@{\hspace{1em}}l@{\hspace{1em}}l@{\hspace{1em}}l@{\hspace{1em}}l@{\hspace{1em}}l@{}}
\textbf{Zheqi Dai\textsuperscript{1,}\thanks{Equal contribution.\quad \textsuperscript{\ensuremath{\dagger}}\,Corresponding authors.}}\kern0.45em &
\textbf{Zhiyue Wu\textsuperscript{2,*}} &
\textbf{Ruibin Yuan\textsuperscript{3}} &
\textbf{Haolin He\textsuperscript{1}} &
\textbf{Chunyat Wu\textsuperscript{1}} &
\textbf{Xuerui Yang\textsuperscript{2}} \\[0.3em]
\multicolumn{6}{@{}l@{}}{%
\textbf{Qiuqiang Kong\textsuperscript{1,\ensuremath{\dagger}}}\hspace{1.5em}%
\textbf{Chao Yan\textsuperscript{2,\ensuremath{\dagger}}}} \\[0.4em]
\multicolumn{6}{@{}l@{}}{\normalfont\footnotesize
$^1$The Chinese University of Hong Kong \quad
$^2$StepFun \quad $^3$MAP}
\end{tabular}}
\title{High-Fidelity Song Generation with\\Joint Discrete and Continuous Modeling}
\author{\authorblock}
\iclrfinalcopy
\begin{document}
\raggedbottom
\sloppy
\maketitle
\lhead{}

\begin{abstract}
Generating a full song from lyrics and a style description requires long-range musical planning without sacrificing vocal and instrumental detail. Low-rate discrete semantic tokens make this tractable, but quantization constrains the information available to acoustic rendering, while separately optimized stages prevent downstream generation objectives from shaping the planner. We propose \textbf{joint discrete and continuous modeling} for high-fidelity song generation. A VQ tokenizer supplies tokens for autoregressive planning, and a separately trained SSL-derived variational encoder supplies continuous semantic targets. Conditioned on the completed token plan and differentiable readouts of the planner's predictive states, a full-sequence flow model generates semantic features for acoustic rendering. Its flow-matching gradients reach the language model through the readouts, without differentiating through token sampling; the standard pathway keeps continuous features out of autoregressive feedback. On 192 shared prompts, its 48D configuration improves SongEval from 4.152 to 4.269 and reduces micro-PER from 20.26\% to 14.47\% relative to a discrete baseline, while receiving substantially higher expert listening scores. We further analyze the system through downstream semantic probes and conditioned reconstruction.
\end{abstract}

\noindent\textbf{Code and demo:} \url{https://dzq84.github.io/JDCM/}

\section{Introduction}
\label{sec:intro}
Generating a full song from lyrics and a style description requires minute-scale organization of vocal melody, accompaniment, and form, as well as detailed acoustic realization. Discrete music representations provide an effective interface between high-level modeling and acoustic synthesis across song generation and music reconstruction \citep{borsos2022audiolm,yang2025songbloom,lei2025levo,dai2026fullsong,xu2024mucodec}. In song generation, a language model predicts low-rate semantic tokens and a downstream model renders audio from their embeddings. This division makes long sequences manageable for the planner while leaving acoustic detail to the renderer.

The economy of a discrete interface also defines its limits. Vector quantization maps distinct self-supervised music features to the same token; looking up that token returns a continuous vector, but cannot recover distinctions lost at assignment. A second limitation concerns training. The planner learns next-token probabilities, whereas a downstream diffusion or flow model learns to synthesize audio. When sampled token IDs are the only connection, the synthesis objective does not update the language model parameters that produced them. The question is therefore how to enrich communication between planning and acoustic rendering without giving up the tractable discrete plan.

DisCon shows that discrete image predictions can guide generation in a continuous space \citep{zheng2025discon}. We bring this principle to music through two aligned representation branches built on an SSL-initialized music encoder: a VQ tokenizer supplies the autoregressive discrete plan, and a separately trained variational encoder supplies continuous semantic targets. Lyric transcription, reconstruction, and teacher alignment shape the latter branch; its features are neither the VQ branch's prequantization features nor its codebook vectors.

Our standard pathway retains discrete tokens as the language model's only autoregressive feedback. Aligned continuous readouts of its causal states, together with the completed token sequence, condition a full-sequence semantic flow; an acoustic renderer converts the generated features into audio. The flow-matching loss updates the planner through the readouts alongside token cross-entropy, without differentiating through hard token sampling. Thus, continuous features inform synthesis and training without entering the language model's autoregressive history.

Feeding generated continuous features back into later planning can introduce an additional exposure bias: subsequent steps use reference features during training but generated features at inference. Our interleaved Joint-SB configuration explores this alternative with feedback in the continuous \emph{semantic} space. The standard pathway avoids this particular mismatch, although the language model still faces a train--inference mismatch in its discrete token history.

On 192 shared WildSongBench prompts, we evaluate six systems with objective quality and style measures, a transcription-derived phoneme-error score, and a listening study in which five music experts assess ten matched song groups. The joint family shows an overall advantage on principal objective and listening measures, while configurations trade off different strengths. MARBLE probes and source-conditioned reconstruction provide intermediate diagnostics alongside full-song evaluation. Our contributions are:

\noindent\textbf{Joint Discrete and Continuous Modeling.} We introduce a jointly trained approach to full-song generation that retains autoregressive planning with discrete semantic tokens while generating continuous semantic features, allowing the flow-matching objective to update the planner; across evaluated configurations, it shows an overall advantage over the discrete baseline on principal objective and expert-listening measures.

\noindent\textbf{Comparing Generation Frameworks Within Our Method.} Within the proposed joint approach, we compare semantic bottleneck widths, target spaces, and generation frameworks across complete systems, including full-sequence semantic synthesis, direct acoustic-latent prediction, and interleaved semantic feedback. Automatic measures of song quality and style alignment, together with expert ratings of sound quality and musicality, reveal trade-offs among these configurations.

\noindent\textbf{Analysis Across Representation and Generation.} We assess discrete and continuous semantic spaces with MARBLE probes, compare source-conditioned reconstruction on 100 in-house songs, and relate these diagnostics to objective and expert-listening results for complete songs. The differing outcomes offer empirical insight into semantic bottlenecks and generation configurations without assuming that reference-conditioned scores predict free generation.

\section{Related Work}
\label{sec:related}
\paragraph{Discrete music representations.}
Jukebox models multiscale VQ-VAE codes autoregressively to produce multi-minute songs, including lyrics-conditioned singing \citep{dhariwal2020jukebox}. AudioLM separates semantic tokens from acoustic codec tokens; MusicLM adopts a semantic--acoustic hierarchy for text-to-music generation, whereas MusicGen autoregressively predicts codec-token streams \citep{borsos2022audiolm,agostinelli2023musiclm,copet2023musicgen}.

\paragraph{Lyrics-conditioned song systems.}
SongGen and YuE extend autoregressive token modeling to songs with lyrics \citep{liu2025songgen,yuan2025yue}. HeartMuLa combines a song language model with a low-frame-rate music codec \citep{yang2026heartmula}. LeVo models mixed and dual-track song tokens, and LeVo~2 uses mixed-token planning states to guide separate vocal and accompaniment token prediction \citep{lei2025levo,lei2026levo2}. Qwen-Music predicts a 25-Hz, single-codebook semantic stream and uses a generative renderer to supply acoustic detail \citep{xu2026qwenmusic}. SongBloom interleaves autoregressive sketches with diffusion refinement and feeds acoustic context into later generation \citep{yang2025songbloom}. Dai et al. use hierarchical token planning before full-sequence flow matching in acoustic VAE latents \citep{dai2026fullsong}.

\paragraph{Diffusion and continuous music generation.}
DiffRhythm generates full songs from lyrics and style with non-autoregressive latent diffusion; DiffRhythm~2 uses semi-autoregressive block flow matching \citep{ning2025diffrhythm,jiang2025diffrhythm2}. ACE-Step combines diffusion with a compressed autoencoder and self-supervised representation alignment \citep{gong2025acestep}. AudioLDM~2 predicts self-supervised audio features to condition latent diffusion, while MusicFlow cascades semantic and acoustic flows \citep{liu2023audioldm2,prajwal2024musicflow}.

\paragraph{Discrete-guided continuous synthesis.}
DisCon is the closest conceptual precedent: a pretrained discrete image model supplies conditions for a separate continuous autoregressive generator, so generation is not restricted to codebook lookups \citep{zheng2025discon}. In our setting, discrete tokens remain explicit prediction targets and also guide a full-sequence semantic flow. The flow additionally receives readouts of the states that predict those tokens, and its loss updates the same language model. Flow-SLM jointly learns speech tokens and continuous acoustic codec frames, with generated frames entering later causal context \citep{chou2025flowslm}. Our standard pathway retains discrete-only autoregressive feedback; the interleaved variant studies feedback in semantic space, unlike the acoustic feedback in SongBloom.

\paragraph{Music tokenization and evaluation.}
MuCodec combines semantic and acoustic information in a quantized music codec, while MuSeC examines the reconstruction--predictability trade-off of music tokens \citep{xu2024mucodec,chen2026musec}. Our continuous target comes from a separate SSL-derived encoder rather than the VQ branch's codebook vectors. MARBLE probes reference-audio representations on music tasks \citep{yuan2023marble}; SongBench and SongEval assess generated songs \citep{wu2026songbench,yao2025songeval}. We use both types of evaluation to distinguish feature utility from the behavior of the full generator.

\section{Joint Discrete and Continuous Modeling}
\label{sec:method}

\begin{figure}[!h]
\centering
\includegraphics[width=\linewidth]{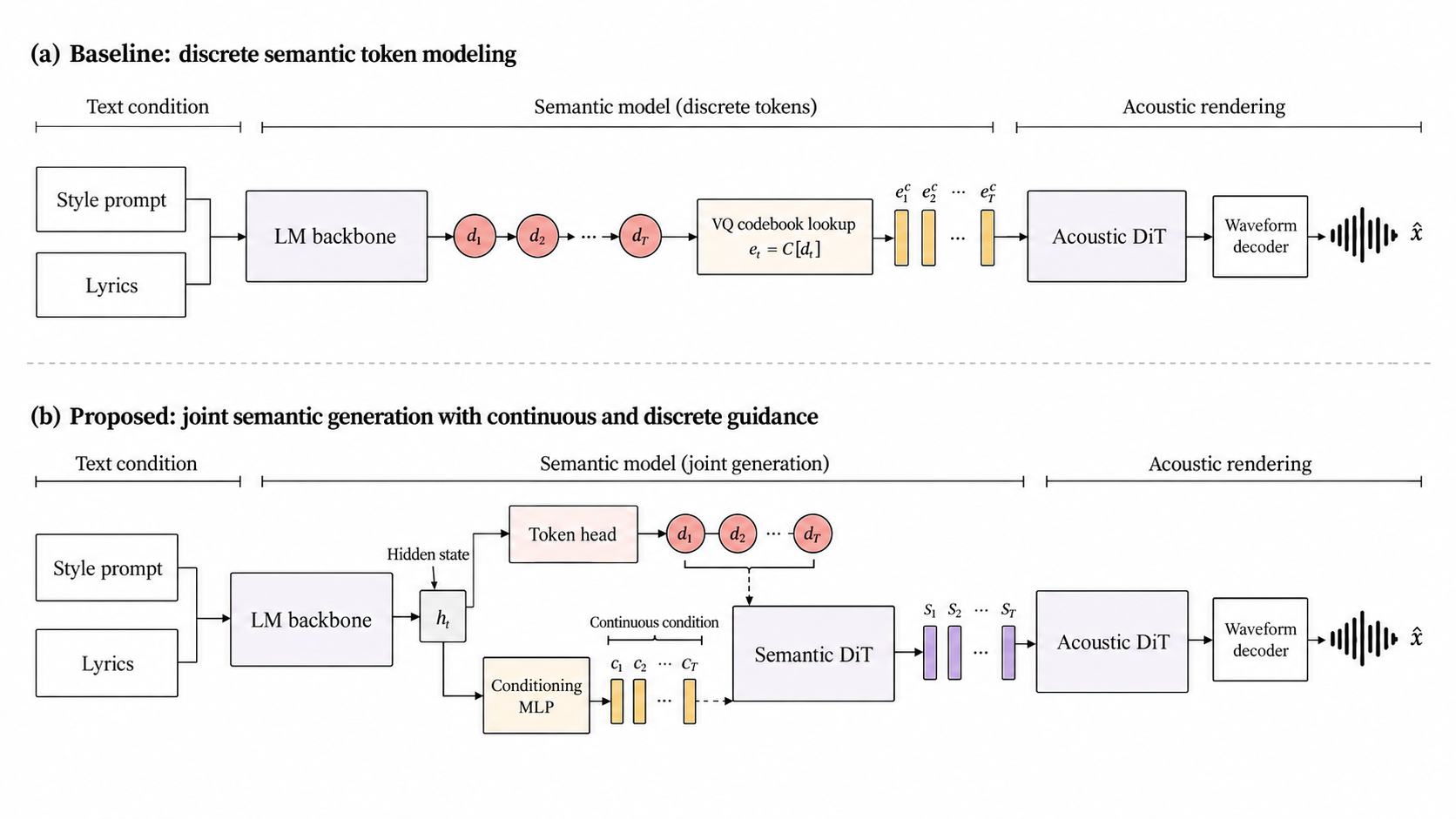}
\caption{Discrete-token song generation and our extension. The baseline conditions acoustic rendering on codebook vectors; our model uses predicted tokens and language-model readouts to generate continuous semantic features. In the schematic, $e_t^c$ is the looked-up vector $e_t$ and $s_t$ is a generated feature $\hat s_t$.}
\label{fig:overview}
\end{figure}

\begin{figure}[!t]
\centering
\makebox[\linewidth][c]{\includegraphics[width=1.08\linewidth]{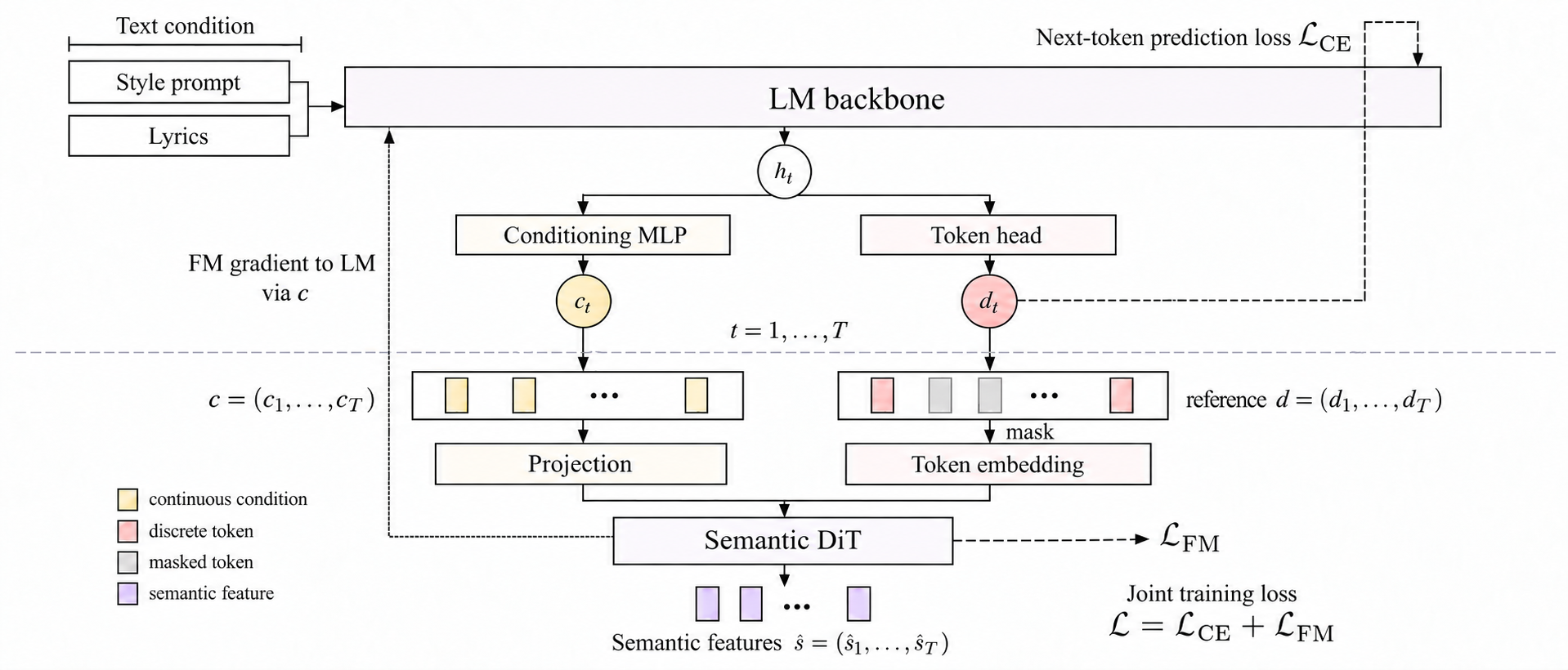}}
\caption{Joint next-token prediction and semantic flow training (Section~\ref*{sec:training}). The token head is supervised to predict $d_t$; the Semantic DiT receives reference IDs with selected positions replaced by an unconditional ID (gray), yielding $\widetilde d$. Its continuous condition $c$ remains intact. Flow-matching gradients reach the language model through $c$; $\hat s$ is a sampled semantic feature sequence, whereas Eq.~\ref*{eq:flow} trains the velocity field.}
\label{fig:joint_training}
\end{figure}

\subsection{Song Generation with Discrete Semantic Tokens}
\label{sec:discrete_interface}

\paragraph{Tokenization and Generation.}
Let $p$ denote the text prompt formed from lyrics and a style description, and let $x$ be a paired reference waveform. The discrete semantic tokenizer extracts low-rate continuous features $u=(u_1,\ldots,u_T)$ from $x$ with an SSL-initialized music encoder \citep{chiu2022bestrq}. A learned projection $\pi$ maps each $u_t$ into the space of a learned VQ codebook $C=\{C_k\}_{k=0}^{K-1}$, where $K$ is the codebook size.\par
\begin{samepage}
\noindent L2 nearest-neighbor assignment \citep{vandenOord2017neural} gives
\begin{equation}
d_t=\arg\min_{0\leq k<K}\|\pi(u_t)-C_k\|_2^2.
\label{eq:tokenizer}
\end{equation}
\end{samepage}
The index $d_t$ is a discrete semantic token; lookup returns a continuous codebook vector $e_t=C_{d_t}$. The tokenizer is detailed in Appendix~\ref{app:representation}.

An autoregressive language model predicts the semantic tokens $d=(d_1,\ldots,d_T)$ given $p$. With a beginning-of-sequence token implicit in the prefix, the probability of these $T$ tokens factors as
\begin{equation}
p_{\rm LM}(d_{1:T}\mid p)
=\prod_{t=1}^{T}p_{\rm LM}(d_t\mid p,d_{<t}).
\label{eq:discrete_lm}
\end{equation}
Audio boundary tokens (BOS and EOS) are included in cross-entropy training; EOS terminates generation. At inference, the language model samples $d$, and codebook lookup gives $e=(e_1,\ldots,e_T)$ to a separately trained acoustic diffusion DiT. Starting from noise, the DiT generates acoustic latents that a decoder turns into a waveform (Figure~\ref{fig:overview}a). The evaluated discrete baseline uses this token-to-codebook interface and a separately trained acoustic renderer.

\paragraph{Strengths and Limitations.}
The low token rate shortens the autoregressive sequence and supports song-level dependencies at practical training and inference cost. Codebook vectors condition the acoustic renderer, which synthesizes sound detail. Such staged systems have produced coherent, high-quality long-form songs.

The interface nevertheless has two limitations. First, vector quantization introduces representation error: distinct encoder features can receive the same token ID, and codebook lookup cannot recover the distinctions discarded by this assignment. The 25-Hz, single-codebook token sequence thus restricts the semantic information passed to the acoustic renderer.

Second, the language model is trained on reference-token cross-entropy, whereas the acoustic DiT is trained separately with a diffusion objective. The acoustic loss therefore does not backpropagate through the language model's predictive states to update its parameters, and token prediction is not jointly optimized for the fidelity of the final audio. Despite the musicality already achieved by this framework, generated songs may retain perceptible audio artifacts.

\subsection{Connecting Discrete Planning to Continuous Semantics}
\label{sec:interface}

We retain discrete tokens as the autoregressive plan and add a readout of the language model's predictive states as a continuous condition (Figure~\ref{fig:overview}b). A semantic flow uses both conditions to generate features from a separately learned continuous target space before acoustic rendering. This design lets the semantic-generation objective update the language model while the flow targets are not limited to codebook vectors.

\paragraph{Continuous semantic tokenizer.}
In the standard configuration, we train an SSL-derived variational encoder $q_\phi$, separately from the discrete tokenizer, that maps the reference waveform $x$ to a distribution over continuous feature sequences $s=(s_1,\ldots,s_T)\in\R^{T\times D}$, where $D$ is the feature width. Each $q_\phi(s_t\mid x)$ is Gaussian with an encoder-predicted mean and diagonal covariance. Samples $s\sim q_\phi(\cdot\mid x)$ are targets for the semantic flow. Their frames align with the discrete tokens $d$, but $s_t$ is neither the prequantization feature $u_t$ nor the codebook vector $e_t$; the discrete tokens are not obtained by quantizing $s$. We freeze this encoder before joint training (Appendix~\ref{app:representation}).

The encoder is trained to retain several aspects of the reference audio in a compact sequence. Lyric-transcription supervision encourages vocal content; mel and chroma reconstruction constrain spectral and harmonic content; alignment encourages agreement with a frozen SSL teacher; and a KL term regularizes the posterior. Together these objectives define the continuous feature space used by the flow. They do not make the representation lossless, nor is the encoder trained to reconstruct a codebook vector. We use posterior samples as flow targets and posterior means for the reference-audio probes, keeping the training target and evaluation feature explicit.

\paragraph{Continuous condition from the language model.}
The conditional factor for $d_t$ in Eq.~\ref{eq:discrete_lm} is parameterized by the state $h_t$ after the language model reads $p$, a beginning-of-sequence symbol, and $d_{<t}$. A two-layer MLP readout $g$ produces
\begin{equation}
c_t=g(h_t).
\label{eq:readout}
\end{equation}
During training, $h_t$ uses ground-truth preceding tokens under teacher forcing; at inference, it uses tokens sampled at earlier steps. After sampling $d_t$, only this discrete token enters the next language-model step: neither $c_t$ nor generated semantic features are fed back in the standard pathway. The aligned sequence $c=(c_1,\ldots,c_T)$ excludes the state that predicts EOS. The readout $c_t$ is a condition for the semantic generator, not the tokenizer output $s_t$ or a direct regression estimate of it.

The semantic DiT embeds the discrete tokens with its own learned table and projects $c$ into its feature space; both condition its velocity prediction from noisy features. The tokens provide explicit guidance, whereas $c_t$ exposes the language model's continuous predictive state to the flow. Because $c_t$ is determined by $p$ and $d_{<t}$, it contains no independent information about a paired reference waveform beyond those inputs. Its contribution is a differentiable condition through which the semantic flow loss can update the planner. The separately learned target $s$ defines the continuous features to be generated; it is not recovered by inverting the discrete quantizer.

Even with the same token prefix, different prompts can produce different readouts. The flow thus receives prompt-dependent planner states alongside the token plan. In the standard pathway, it processes the completed sequences bidirectionally after autoregressive planning. Continuous features do not enter the language model's history; the planner still sees sampled rather than reference token prefixes at inference.

The semantic DiT's token-embedding table is distinct from the tokenizer codebook $C$. In the standard configuration, we initialize its continuous-condition adapter to zero, leaving the pretrained token-conditioned flow's initial prediction unchanged. Training configurations are in Appendix~\ref{app:training}.

\subsection{Joint Training}
\label{sec:training}

In the standard configuration, for each training pair $(p,x)$, the fixed discrete tokenizer supplies reference tokens $d$, and the frozen continuous tokenizer supplies a sample $s\sim q_\phi(\cdot\mid x)$. As Figure~\ref{fig:joint_training} shows, the language model retains teacher-forced token cross-entropy $\mathcal L_{\rm CE}$, including sequence boundaries. For semantic flow matching, draw standard Gaussian noise $\epsilon\in\R^{T\times D}$ and time $\tau\sim\mathcal U[0,1]$. The interpolated state $s_\tau=(1-\tau)\epsilon+\tau s$ has target velocity $s-\epsilon$. Let $\widetilde d$ denote the DiT's token condition obtained from reference $d$. During training, independently selected positions are replaced by a dedicated unconditional ID (gray entries in Figure~\ref{fig:joint_training}); a separate example-level dropout replaces the entire sequence for selected examples. Neither operation changes the language model's reference-token prefix or its readout $c$. The semantic DiT predicts the velocity with $v_\theta$, giving the flow loss for an unpadded sequence:
\begin{equation}
\mathcal L_{\rm FM}
=\E\!\left[\frac{1}{TD}
\left\|v_\theta(s_\tau,\tau;\widetilde d,c)-(s-\epsilon)\right\|_F^2\right].
\label{eq:flow}
\end{equation}
The expectation covers training pairs, posterior samples, noise, time, and token-condition corruption. We minimize $\mathcal L_{\rm CE}+\mathcal L_{\rm FM}$ with equal weights. The flow loss reaches the unfrozen language-model layers through the continuous readout and its adapter, while the token loss directly supervises autoregressive prediction; neither path requires a gradient through hard token assignments. Both token-condition perturbations are disabled at inference. Both tokenizers and the downstream acoustic renderer remain fixed. The padded-batch flow loss is given in Appendix~\ref{app:training}.

At initialization, the zero-initialized adapter preserves the pretrained token-conditioned flow. Its first update trains the adapter; subsequent flow gradients can reach the readout and unfrozen language-model layers. Token-condition masking is intended to reduce exclusive reliance on token identities. The present comparisons do not isolate its contribution.

\subsection{Full-Song Generation}
\label{sec:generation}

In the standard configuration, given a new prompt $p$, the language model samples discrete tokens until EOS while retaining their aligned readouts $c$. The semantic flow generates a complete continuous sequence from noise, conditioned on $(d,c)$. A separately trained acoustic diffusion renderer converts this sequence into acoustic latents, which a decoder turns into the output waveform (Appendix~\ref{app:acoustic}). The continuous tokenizer is used to provide training targets, not at inference. 

In the standard pathway, planning proceeds at 25 semantic frames per second. Once tokens are complete, each semantic flow step updates the full feature sequence in parallel. The acoustic model then renders windows and decodes their latents. End-to-end latency depends on all three stages; it cannot be inferred from token rate alone.

\section{Experiments}
\label{sec:experiments}
\subsection{Experimental setup}
\label{sec:setup}
\paragraph{Systems and training.}
Six complete song-generation systems are trained on in-house synthetic music data and share a LeVo-VAE waveform decoder \citep{lei2025levo}. The standard joint pathway uses a Qwen2.5-1.5B language-model backbone \citep{qwen2024qwen25}. Discrete sends VQ lookups to acoustic diffusion. Joint-STD-48D, Joint-STD-96D, and Joint-STD-128D pair token planning with full-sequence semantic flow and width-specific acoustic renderers. Joint-LeVo-64D generates acoustic latents directly by full-sequence flow, without a second acoustic DiT. Joint-SB-48D interleaves token planning and semantic-patch flow, feeding prior \emph{semantic} patches back to the planner, unlike SongBloom's acoustic feedback \citep{yang2025songbloom}. The LeVo and SB labels denote internal configurations. Joint training updates the planner, readout, and DiT flow while keeping the tokenizers and renderer fixed.

\paragraph{Evaluation and measures.}
Each system yields 192 retained songs on the same WildSongBench prompts \citep{map2026wildsongbench}, for 1,152 outputs. The standard worker retries token sequences shorter than 30 seconds in fixed seed order, retaining the first sufficiently long result or the longest fallback; Joint-SB uses its own sampler. Shared prompts and scoring procedures permit system-level comparison. We report SongBench and SongEval means \citep{wu2026songbench,yao2025songeval}, AudioBox production quality \citep{tjandra2025audiobox}, and MuQ--MuLan style similarity \citep{zhu2025muq}. For subjective evaluation, five music experts rate ten matched song groups on sound quality and musicality using 1--5 scales; we report the mean ratings as MOS. Micro-PER pools phoneme edits after choosing the lowest-error of four Qwen3-ASR passes \citep{shi2026qwen3asr} on each saved song. MARBLE probes \citep{yuan2023marble} on reference-audio features are analyzed separately in Section~\ref{sec:probes}.

\subsection{Song generation results}
\label{sec:mainresults}
Table~\ref{tab:main} compares the six complete systems. Relative to Discrete, Joint-STD-48D improves SongBench (6.282 versus 6.217), SongEval (4.269 versus 4.152), and style alignment (0.514 versus 0.487), while reducing micro-PER from 20.260\% to 14.470\%. Its production-quality score is also slightly higher (8.300 versus 8.282), and the expert MOS scores favor it on both listening dimensions.

\begin{table}[!h]
\centering\small
\setlength{\tabcolsep}{3.5pt}
\caption{Six systems on 192 WildSongBench prompts. PER is selected-transcript micro phoneme error (\%). RTF is generation wall-clock time divided by audio duration. MOS denotes mean ratings by five music experts on ten matched song groups (sound quality and musicality; 1--5). Across all six systems, bold marks the best and underlining marks the second-best value in each column; tied second-best values are all underlined.}
\label{tab:main}
\resizebox{\linewidth}{!}{%
\begin{tabular}{lrrrrrrrr}
\toprule
System & SongBench$\uparrow$ & $\mathrm{PQ}_{\rm AudioBox}\uparrow$ & SongEval$\uparrow$ & $\mathrm{MuQ}_{\rm MuLan}\uparrow$ & PER$\downarrow$ & RTF$\downarrow$ & $\mathrm{MOS}_{\rm quality}\uparrow$ & $\mathrm{MOS}_{\rm music}\uparrow$\\
\midrule
Discrete & 6.217 & 8.282 & 4.152 & 0.487 & 20.26 & \textbf{1.511} & 3.14 & 2.76\\
\midrule
Joint-STD-48D & \underline{6.282} & \underline{8.300} & \underline{4.269} & \underline{0.514} & \textbf{14.47} & 1.879 & \textbf{4.14} & \textbf{3.26}\\
Joint-STD-96D & 6.263 & \textbf{8.301} & 4.222 & 0.503 & \underline{17.25} & 1.779 & 3.38 & \underline{3.22}\\
Joint-STD-128D & 6.251 & 8.276 & 4.145 & \textbf{0.518} & 18.86 & \underline{1.682} & \underline{3.76} & 3.14\\
\midrule
Joint-LeVo-64D & 6.207 & 8.281 & 4.230 & 0.507 & 18.02 & 1.728 & 3.04 & 3.10\\
Joint-SB-48D & \textbf{6.311} & 8.282 & \textbf{4.288} & 0.495 & 18.21 & 2.434 & 3.38 & 2.88\\
\bottomrule
\end{tabular}%
}
\end{table}

RTF describes the observed cost of each complete generation pipeline under the current evaluation conditions.

\begin{figure}[!h]
\centering
\includegraphics[width=0.96\linewidth]{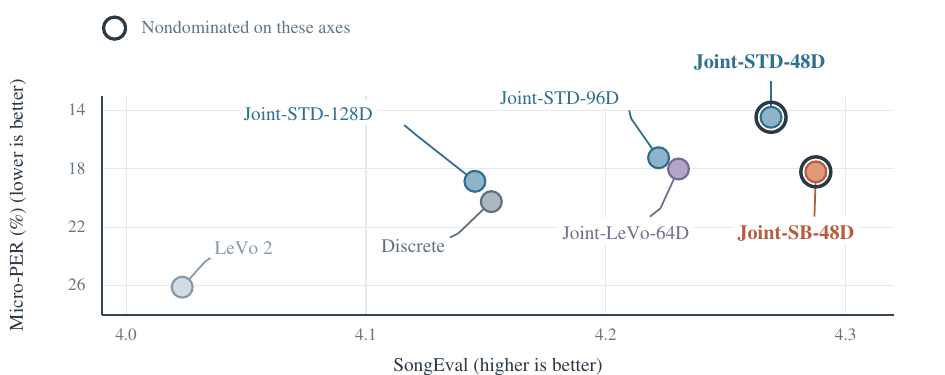}
\caption{SongEval versus micro-PER for six evaluated systems and a published reference \citep{map2026wildsongbench}. Outlined points are nondominated among the six systems. The external reference follows its source candidate-selection protocol.}
\label{fig:wsb_tradeoff}
\end{figure}

Figure~\ref{fig:wsb_tradeoff} shows the SongEval--PER trade-off: Joint-STD-48D and Joint-SB-48D are the two nondominated internal systems on the plotted axes. Within the standard pathway, increasing target width from 48D to 96D and 128D does not improve SongEval or PER; 96D marginally leads production quality, while 128D leads style alignment. Joint-LeVo-64D explores direct acoustic-latent prediction, and Joint-SB-48D explores semantic feedback.

\subsection{Downstream utility of semantic representations}
\label{sec:probes}
\begin{table}[!h]
\centering\small
\setlength{\tabcolsep}{3.5pt}
\caption{MARBLE probes on reference-audio features. Column labels give the dataset or task, with the metric as a subscript: EMO is EmoMusic and GTZAN-R is beat tracking on GTZAN Rhythm; A and V denote arousal and valence. The original SSL row is a separate reference; bold and underlining mark the best and second-best compact representations. VQ uses looked-up vectors, and continuous rows use unprojected bottleneck features.}
\label{tab:probes}
\begin{tabular}{lrrrrrr}
\toprule
Representation & GTZAN$_{\rm Acc.}$ & EMO$_{{\rm A},R^2}$ & EMO$_{{\rm V},R^2}$ & GTZAN-R$_{F_1}$ & Downbeat$_{F_1}$ & Structure$_{\rm Acc.}$\\
\midrule
Original SSL-1024D & 0.703 & 0.648 & 0.439 & 0.657 & 0.449 & 0.553\\
\midrule
VQ lookup & 0.459 & 0.447 & 0.143 & \textbf{0.328} & 0.107 & 0.483\\
Continuous-48D & 0.472 & 0.514 & 0.223 & 0.323 & \textbf{0.110} & \underline{0.514}\\
Continuous-96D & \underline{0.510} & \underline{0.541} & \underline{0.250} & \underline{0.325} & 0.107 & 0.510\\
Continuous-128D & \textbf{0.541} & \textbf{0.557} & \textbf{0.297} & 0.317 & \underline{0.108} & \textbf{0.520}\\
\bottomrule
\end{tabular}
\end{table}
Table~\ref{tab:probes} separates the original 1024D SSL reference from compact representations. Among the listed compact rows, 128D leads genre, arousal, valence, and structure; VQ leads beat, while continuous 48D leads downbeat. The 48D generator nevertheless has the lowest PER in Table~\ref{tab:main}. Relative to VQ lookups, its reference features improve five of the six probes, with beat F1 the exception.

\subsection{Reference-conditioned reconstruction}
\label{sec:condition_diagnostic}
Table~\ref{tab:reconstruction_selected} reports reconstruction diagnostics on 100 in-house songs under two source-conditioned protocols. The original audio is scored by the same procedure as a reference.

\begin{table}[!h]
\centering\small
\setlength{\tabcolsep}{8pt}
\caption{Reconstruction on 100 in-house songs (lower is better). Mel Mag and Chroma Mag are magnitude errors; CTC is the transcription loss and CER is character error rate. The original audio is a scoring reference. Bold and underlining mark the best and second-best system within each reconstruction protocol.}
\label{tab:reconstruction_selected}
\begin{tabular}{lrrrr}
\toprule
System & Mel Mag$\downarrow$ & Chroma Mag$\downarrow$ & CTC$\downarrow$ & CER$\downarrow$\\
\midrule
Original audio & 0.000 & 0.000 & 1.033 & 18.52\%\\
\midrule
\multicolumn{5}{@{}l}{\emph{Ground-truth acoustic-latent reconstruction}}\\
Discrete & 0.392 & 0.198 & 2.007 & 41.41\%\\
Joint-STD-48D & \underline{0.273} & \underline{0.096} & \underline{1.331} & \underline{25.93\%}\\
Joint-STD-96D & 0.287 & 0.113 & 1.454 & 30.30\%\\
Joint-STD-128D & 0.296 & 0.126 & 1.465 & 28.62\%\\
Joint-LeVo-64D & \textbf{0.256} & \textbf{0.078} & 1.357 & 26.94\%\\
Joint-SB-48D & 0.275 & 0.098 & \textbf{1.223} & \textbf{22.90\%}\\
\midrule
\multicolumn{5}{@{}l}{\emph{Teacher-forced reconstruction}}\\
Discrete & 0.450 & 0.248 & 1.746 & 31.31\%\\
Joint-STD-48D & \underline{0.440} & \underline{0.247} & \underline{1.586} & \underline{30.64\%}\\
Joint-STD-96D & 0.488 & 0.300 & 2.008 & 42.76\%\\
Joint-STD-128D & 0.486 & 0.275 & 1.811 & 36.36\%\\
Joint-LeVo-64D & 0.495 & 0.284 & 2.094 & 40.07\%\\
Joint-SB-48D & \textbf{0.362} & \textbf{0.205} & \textbf{1.348} & \textbf{28.28\%}\\
\bottomrule
\end{tabular}
\end{table}

Under ground-truth acoustic latents, Joint-LeVo-64D has the lowest Mel and Chroma magnitude errors among the systems, while Joint-SB-48D has the lowest CTC and CER. Joint-SB-48D leads all four reported measures under teacher forcing.

\section{Discussion and Limitations}
\label{sec:discussion}
\paragraph{System-level gains and their attribution.}
The joint pathway retains a low-rate discrete plan while exposing the language model's predictive states to semantic synthesis through a differentiable readout. Relative to Discrete, Joint-STD-48D reduces micro-PER by 5.79 percentage points (20.26\% to 14.47\%), raises SongEval from 4.152 to 4.269, and raises the reported sound-quality MOS from 3.14 to 4.14 (Table~\ref{tab:main}). AudioBox PQ changes by only 0.018. These measures emphasize different properties of the same songs; a single quality score would obscure the concurrent changes in lyric fidelity and listening judgments. The comparison supports the viability of the complete joint system, but cannot assign its gains specifically to the flow gradient, the continuous target, or the renderer, all of which differ from the discrete baseline. The readout also cannot recover details lost from a particular reference recording: it is determined by the prompt and token history.

\paragraph{Semantic capacity and generation behavior.}
Larger reference-audio features do not yield uniformly stronger generation. From 48D to 128D, genre accuracy rises from 0.472 to 0.541 and valence $R^2$ from 0.223 to 0.297 in Table~\ref{tab:probes}; across the corresponding complete systems, SongEval falls from 4.269 to 4.145 and micro-PER rises from 14.47\% to 18.86\% in Table~\ref{tab:main}. A richer feature space may be harder to predict or render, but these comparisons cannot establish that mechanism because the training paths and renderers are not fully matched. Figure~\ref{fig:wsb_tradeoff} shows a separate trade-off in the generation schedule: Joint-SB-48D has the highest SongBench and SongEval point estimates (6.311 and 4.288), whereas Joint-STD-48D has the lowest micro-PER (14.47\% versus 18.21\%). Semantic feedback is therefore an informative design alternative, although these complete-system results do not isolate its effect.

\paragraph{What reconstruction can diagnose.}
The 100-song source-conditioned study in Table~\ref{tab:reconstruction_selected} probes a different failure mode from free generation. With ground-truth acoustic latents, Joint-LeVo-64D has the lowest Mel and Chroma magnitude errors (0.256 and 0.078), yet its free-generation SongBench score is 6.207, below Joint-STD-48D's 6.282. Under teacher forcing, Joint-SB-48D leads all four reconstruction measures, while its free-generation lyric error remains higher than Joint-STD-48D's. Access to reference conditions can expose what a renderer preserves, but cannot show how reliably the planner supplies those conditions. Even the original audio has nonzero transcription loss and CER in this diagnostic, so transcription-derived measures should be read with their measurement process in mind.

\paragraph{Evaluation and remaining controls.}
The automatic results are point estimates from one retained song per system and prompt; the five-expert MOS study covers ten matched groups rather than all 192 prompts. The 0.001 AudioBox PQ difference between Joint-STD-96D and Joint-STD-48D, for example, is too small to interpret without paired uncertainty. Micro-PER selects the lowest-error result of four Qwen3-ASR passes, and the standard worker's duration retries differ from Joint-SB's sampling protocol. These choices limit direct comparison with other candidate-selection rules and do not remove recognition or text-normalization error. Matched variants that block the flow gradient, remove the readout, or directly reconstruct semantic features from language-model states would identify the contribution of each connection under the same target, renderer, and training budget.

\paragraph{Inference mismatch and cost.}
The standard pathway keeps generated continuous features out of the language model's autoregressive history, avoiding semantic-feature feedback mismatch. It still learns from reference token prefixes and generates from sampled ones; the acoustic renderer likewise encounters predicted rather than reference semantic features at inference. Interleaved semantic feedback adds another train--inference mismatch to examine. Full-sequence semantic flow also waits for the completed token plan, and the additional representation and acoustic models incur training and inference cost. The present experiments establish neither a compute advantage over direct acoustic generation nor an optimal width across datasets, languages, and budgets.

\section{Conclusion}
Joint discrete and continuous modeling extends token-guided song generation with a continuous condition and joint language-model and semantic-flow training. Discrete tokens remain the autoregressive history and guide synthesis, while semantic generation uses the language model's predictive states through a learned readout. On a common 192-prompt evaluation, Joint-STD-48D improves the main automatic metrics over the discrete baseline; macro-level bottleneck and architecture ablations show different quality and style-alignment outcomes. Representation probes further show that no compact target width leads every measured task. Finer component ablations remain a direction for future work.

\clearpage
\subsection*{AI use statement}
Generative AI assisted with polishing. The authors are responsible for verifying the AI-assisted text and claims and for the final content of the paper.
\subsection*{Ethics statement}
Song synthesis can reproduce recognizable vocal characteristics and musical material. Generated audio and automatic assessments should not be presented as recordings or judgments from real performers or listeners.
\subsection*{Reproducibility statement}
The main text specifies the representation paths, joint-training objective, and generation systems; the appendix details the representation targets, semantic DiT pretraining, joint semantic training, and acoustic rendering. The code and demo address is given below the abstract.
\bibliography{references}
\bibliographystyle{iclr2027_conference}
\clearpage
\appendix
\numberwithin{table}{section}
\section{Semantic Representation Targets}
\label{app:representation}

We use three separately trained representation paths: the original 1024D SSL
encoder for MARBLE probing, a single-codebook VQ tokenizer for discrete semantic
tokens, and an SSL-VAE for continuous semantic targets. All representations use
24-kHz audio and operate at 25~Hz. The discrete tokenizer adopts a multi-stage
representation-learning and tokenization paradigm similar to prior music
tokenization systems~\citep{stepaudio3music,xu2026qwenmusic}, consisting of
BEST-RQ-style self-supervised pretraining, CTC/Mel/Chroma supervision, and
single-codebook vector quantization. The SSL-VAE is initialized from a separate
Stage-2 checkpoint and introduces a Gaussian bottleneck for continuous semantic
modeling. The three representation paths use separate weights.

For each frame, the SSL-VAE defines a diagonal Gaussian posterior and samples
the continuous semantic target through reparameterization:
\begin{equation}
q(s_t|x)=\mathcal{N}\!\left(\mu_t,\mathrm{diag}(e^{\nu_t})\right),
\qquad
s_t=\mu_t+\exp(\nu_t/2)\odot\xi_t,\quad
\xi_t\sim\mathcal{N}(0,I).
\label{eq:sslvae}
\end{equation}
Posterior samples $s_t$ are used as semantic-flow training targets, whereas
posterior means $\mu_t$ are used for deterministic representation probing.

The SSL-VAE is optimized with
\begin{equation}
\mathcal{L}_{\mathrm{rep}}
=
2\mathcal{L}_{\mathrm{CTC}}
+\mathcal{L}_{\mathrm{mel}}
+\mathcal{L}_{\mathrm{chr}}
+10^{-5}\mathcal{L}_{\mathrm{KL}}
+\mathcal{L}_{\mathrm{align}}.
\label{eq:repr-loss}
\end{equation}
For $r\in\{\mathrm{mel},\mathrm{chr}\}$, the spectral reconstruction loss is
the sum of spectral convergence and magnitude $L_1$ losses:
\begin{equation}
\mathcal{L}_{r}
=
\frac{\|Y_r-\hat{Y}_r\|_F}{\|Y_r\|_F}
+
\frac{1}{N_r}\|Y_r-\hat{Y}_r\|_1,
\label{eq:spec-loss}
\end{equation}
where $Y_r$ and $\hat{Y}_r$ denote the reference and predicted Mel or Chroma
features, respectively, and $N_r$ is the number of feature elements.
Key representation configurations are summarized in
Table~\ref{tab:repr-config}.

\begin{table}[h]
\caption{\textbf{Semantic representation configuration}. Both representation
paths use 24-kHz audio, a 24-block non-causal Conformer, and temporally aligned
25-Hz features.}
\label{tab:repr-config}
\centering
\omctablefont
\setlength{\tabcolsep}{6pt}
\begin{tabular*}{\textwidth}{@{\extracolsep{\fill}}lcc@{}}
\toprule
\textbf{Setting} & \textbf{Discrete Tokenizer} & \textbf{Continuous SSL-VAE} \\
\midrule
Bottleneck Position & After block 12 & After block 12 \\
Bottleneck Type & Single-codebook VQ & Diagonal Gaussian \\
Codebook Size & 32,768 & -- \\
Training Objectives & CTC + Mel + Chroma & CTC + Mel + Chroma + KL + Align \\
Training Target & Discrete token $d_t$ & Posterior sample $s_t$ \\
Evaluation Feature & Codebook vector $C[d_t]$ & Posterior mean $\mu_t$ \\
\bottomrule
\end{tabular*}
\end{table}

\section{Joint Semantic Generator and Training}
\label{app:training}

\paragraph{Discrete LM pretraining.}
The discrete song planner is initialized from Qwen2.5-1.5B \citep{qwen2024qwen25} and trained as a
text-conditioned autoregressive music-token model. Each training sequence
consists of a formatted lyric-and-style text prefix followed by audio BOS,
25-Hz semantic tokens, and EOS. The complete text prefix is visible to all
music-token positions but is excluded from the training loss; next-token
cross-entropy is applied only to the audio-token sequence, including the audio
boundaries. The learning rate is linearly warmed up for 2,000 steps to
$3\times10^{-4}$, kept constant for most of training, and linearly annealed
to zero over the final 10\% of training steps.

\paragraph{Standalone semantic DiT pretraining.}
Before joint training, the semantic DiT is trained to generate SSL-VAE features
from noise using reference semantic IDs alone. This ID-conditioned model
initializes the semantic DiT in the joint pathway; the language-model readout
is added only during joint training. Table~\ref{tab:semantic-dit-warmup}
summarizes the standalone configuration.

\begin{table}[!htbp]
\caption{\textbf{Standalone semantic DiT pretraining.} The model learns
SSL-VAE features conditioned on reference semantic IDs before joint semantic
training.}
\label{tab:semantic-dit-warmup}
\centering
\omctablefont
\setlength{\tabcolsep}{4pt}
\begin{tabular*}{\textwidth}{
@{\extracolsep{\fill}}
>{\raggedright\arraybackslash}p{0.25\textwidth}
>{\raggedright\arraybackslash}p{0.67\textwidth}
@{}}
\toprule
\textbf{Setting} & \textbf{Semantic DiT Warm-up} \\
\midrule
Architecture & 1D DiT-XL; 28 blocks, width 1152, 16 heads \\
Conditioning & Ground-truth 25-Hz semantic IDs; no LM hidden-state readout \\
Prediction Target & SSL-VAE posterior features, $D\in\{48,96,128\}$ \\
Objective & Flow-matching velocity MSE, $\|v_\theta(s_\tau,\tau;d)-(s-\epsilon)\|_F^2/(TD)$ \\
Flow Path & $s_\tau=(1-\tau)\epsilon+\tau s$, with $\epsilon\sim\mathcal N(0,I)$ \\
Sequence & Frame-aligned full song at 25 Hz; maximum 8192 frames \\
Trainable & Semantic DiT only \\
Frozen & Semantic tokenizer and SSL-VAE target encoder; LM not used \\
LR Warmup & 5,000 steps \\
Peak LR & $5\times10^{-5}$ \\
LR Schedule & Warmup followed by cosine decay to $10^{-7}$ at step 500,000 \\
Sampling & 32 Euler flow steps \\
\bottomrule
\end{tabular*}
\end{table}

\newpage
\paragraph{Joint semantic training.}
The standard joint pathway uses the pretrained song LM with a separate
audio-token embedding and prediction head. A two-layer readout maps the
language model's predictive states to continuous conditions $c_t$ for a
semantic DiT. The semantic DiT predicts continuous SSL-VAE features from
noise, conditioned jointly on discrete token embeddings and the continuous
readouts. The continuous tokenizer and downstream acoustic renderer remain
fixed during joint training.

The readout $c_t$ is aligned with the token predicted by the corresponding
language-model state: the BOS state predicts the first semantic frame, and
the state after $d_{T-1}$ predicts the final frame. The state predicting EOS
has no aligned semantic target. Training uses reference-token prefixes,
whereas inference uses sampled prefixes.

The joint objective is
\begin{equation}
\mathcal{L}_{\mathrm{joint}}
=
\mathcal{L}_{\mathrm{CE}}
+
\mathcal{L}_{\mathrm{FM}},
\label{eq:joint-loss-app}
\end{equation}
where token cross-entropy supervises autoregressive semantic-token prediction
and the flow-matching loss trains the semantic DiT. For a padded minibatch,
the implemented flow loss is
\begin{equation}
\ell_{\mathrm{FM}}
=
\frac{
\sum_{i,t} m_{it}
\|v_{it}-(s_{it}-\epsilon_{it})\|_2^2 / D
}{
\sum_{i,t}m_{it}
},
\label{eq:fm-loss-app}
\end{equation}
where $m_{it}$ marks valid semantic frames, $s$ is sampled from the frozen
SSL-VAE posterior, and $\epsilon$ is the flow noise. Token corruption is
applied only to the semantic DiT condition and does not modify the
teacher-forced language-model history. The continuous readout is not detached,
allowing flow-matching gradients to update the language model without
differentiating through discrete token selection.

Key configurations are summarized in Table~\ref{tab:joint-config}.

\begin{center}
\begin{minipage}{\textwidth}
\centering
\omctablefont
\makeatletter\def\@captype{table}\makeatother
\caption{\textbf{Discrete planner pretraining and joint semantic training configuration}.
The discrete LM is first trained with text-conditioned next-token prediction.
Joint training subsequently optimizes the pretrained planner, continuous readout,
and semantic DiT while keeping the representation encoders and acoustic renderer fixed.}
\label{tab:joint-config}
\setlength{\tabcolsep}{4pt}
\begin{tabular*}{\textwidth}{
@{\extracolsep{\fill}}
>{\raggedright\arraybackslash}p{0.20\textwidth}
>{\raggedright\arraybackslash}p{0.31\textwidth}
>{\raggedright\arraybackslash}p{0.41\textwidth}
@{}}
\toprule
\textbf{Setting}
& \textbf{Discrete LM Pretraining}
& \textbf{Joint Semantic Training}
\\
\midrule

Base Model
& Qwen2.5-1.5B
& Pretrained Music LM + Semantic DiT
\\

Conditioning
& Lyrics + style
& IDs + LM readout
\\

Prediction Target
& 25-Hz semantic IDs
& SSL-VAE features
\\

Objective
& Audio-token NTP
& $\mathcal{L}_{\mathrm{CE}}+\mathcal{L}_{\mathrm{FM}}$
\\

Sequence
& Text + BOS + IDs + EOS
& Frame-aligned conditions
\\

Trainable
& LM
& LM + readout + DiT
\\

Frozen
& Semantic tokenizer
& Tokenizers + acoustic renderer
\\

LR Warmup
& 2,000 steps
& None
\\

Peak LR
& $3\times10^{-4}$
& DiT + latent head: $5\times10^{-5}$; LM: $1\times10^{-5}$
\\

LR Schedule
& Constant; final 10\% decay to 0
& Constant; no scheduler
\\

\bottomrule
\end{tabular*}
\end{minipage}
\end{center}

\section{Acoustic Rendering}
\label{app:acoustic}

For the standard SSL-VAE pathway, a separately trained acoustic DiT maps
semantic features to LeVo VAE acoustic latents. It receives reference SSL-VAE
features during its own training and generated semantic features at song
generation time. The acoustic renderer is frozen during joint semantic
training. Table~\ref{tab:acoustic-dit-selected} reports the selected
width-specific renderers.

\begin{table}[!htbp]
\caption{\textbf{Selected acoustic DiT renderers for the standard SSL-VAE route.}
Each representation width uses a separate acoustic checkpoint.
$^{\dagger}$The 96D/128D learning-rate values come from the selected checkpoint
directory names; the original trainer configuration was not recovered.}
\label{tab:acoustic-dit-selected}
\centering
\omctablefont
\setlength{\tabcolsep}{4pt}
\begin{tabular*}{\textwidth}{
@{\extracolsep{\fill}}
>{\raggedright\arraybackslash}p{0.25\textwidth}
>{\raggedright\arraybackslash}p{0.67\textwidth}
@{}}
\toprule
\textbf{Setting} & \textbf{Selected Acoustic DiT Renderers} \\
\midrule
Architecture & Continuous DiT; 24 blocks, width 1536, 24 heads \\
Conditioning & Reference SSL-VAE features ($D=48,96,128$); preceding 50 acoustic latent frames as context; 128D uses condition LayerNorm \\
Prediction Target & 64-channel LeVo VAE acoustic latents \\
Objective & Trigonometric diffusion $v$-prediction \\
Sequence & 25-Hz, 750-frame (30-s) windows with 50-frame (2-s) context \\
Trainable & Acoustic DiT; trained separately and frozen during joint semantic training \\
Frozen & SSL-VAE feature extractor and pretrained LeVo VAE decoder \\
LR Warmup & 5,000 steps \\
Peak LR & $5\times10^{-5}$ \\
LR Schedule & Warmup followed by cosine decay to $10^{-7}$ at step 500,000 \\
Sampling & 32 deterministic $v$-prediction updates, then 48-kHz stereo LeVo decoding \\
\bottomrule
\end{tabular*}
\end{table}

\end{document}